\documentclass[api,prl,showpacs,letterpaper,twocolumn,groupedaddress,footinbib,amssymb]{revtex4}

\usepackage{graphicx}

\begin{document}

\title{Intriguing magnetism of Fe monolayers on hexagonal transition-metal
surfaces}

\author{B.~Hardrat$^{a}$}
\author{A.~Al-Zubi$^{b}$}
\author{P.~Ferriani$^{a}$}
\author{S.~Bl\"ugel$^{b}$}
\author{G.~Bihlmayer$^{b}$}
\author{S.~Heinze$^{a}$}
\email[corresp.\ author: ]{heinze@physnet.uni-hamburg.de}

\affiliation{$^{a}$Institute of Applied Physics, University of
Hamburg, Jungiusstr.~11, 20355 Hamburg, Germany}
\affiliation{$^{b}$Institut f\"{u}r Festk\"{o}rperforschung,
             Forschungszentrum J\"{u}lich, D-52425 J\"{u}lich, Germany}

\date{\today}

\begin{abstract}
Using first-principles calculations, we demonstrate that an Fe
monolayer can assume very different magnetic phases on hexagonal
hcp (0001) and fcc (111) surfaces of $4d$- and $5d$-transition
metals. Due to the substrates' $d$-band filling, the 
nearest-neighbor exchange coupling of Fe changes gradually from
antiferromagnetic (AFM) for Fe films on Tc, Re, Ru and Os to
ferromagnetic
on Rh, Ir, Pd, and Pt. In combination with the topological
frustration on the triangular lattice of these surfaces the AFM
coupling results in a 120$^\circ$ N\'eel structure for Fe on
Re and Ru and an unexpected double-row-wise AFM
structure on Rh, which is a superposition of a left- and
right-rotating 90$^\circ$ spin spiral.
\end{abstract}

\maketitle

Triggered by the discovery of the giant-magnetoresistance effect
and to realize novel spintronic device concepts~\cite{Fert2007},
magnetic nanostructures on surfaces have been a focus of
experimental and theoretical research for more than 20 years now.
In particular, there has been a tremendous effort to grow
ultra-thin transition-metal films on metal surfaces and to
characterize and explain their magnetic properties. It is now
generally believed that these structurally simple systems are well
understood and more complex nanostructures such as atomic chains,
clusters, or molecules on surfaces have moved into the spotlight
of today's research.

Therefore, it came as a big surprise when it was experimentally
shown that the prototypical ferromagnet Fe becomes a
two-dimensional (2D) antiferromagnet on the W(001)
surface~\cite{KubetzkaPRL2005}.
That complex magnetic order can be obtained even in
single monolayer (ML) magnetic films on non-magnetic
substrates has been dramatically
demonstrated by the recent discovery of a spin-spiral state for a
Mn ML on W(110)~\cite{Bode2007} and a Mn ML on
W(001)~\cite{Ferriani2008} and a nanoscale magnetic structure for
an Fe ML on Ir(111)~\cite{vBergmannPRL2006}.
Surfaces of $4d$- and $5d$-transition metals (TMs) such as W, Re,
Ru, or Ir have been particularly attractive from an experimental 
point of view as ultra-thin $3d$-TM films can often be grown
pseudomorphically and without
intermixing~\cite{AlbrechtPRL2000,GabalyPRL2006,LiuPRB1990,AndrieuPRB1992,RepettoPRB2006}.
However, there has been a controversy in the past about reports  
concerning dead magnetic layers and absence of magnetic order in
ultra-thin films on these
surfaces~\cite{LiuPRB1990,AndrieuPRB1992}. The fundamental key to
many unresolved puzzles may be the itinerant character of TMs
resulting in competing exchange interactions beyond
nearest-neighbors and higher-order spin interactions beyond the
Heisenberg model. The latter interactions have been
proposed to play a role in transition-metals, however, to our
knowledge no unambiguous
proof of their importance has been given.

Here, we use first-principles calculations to demonstrate that an
Fe ML can assume very different magnetic phases on
a triangular lattice provided by hcp (0001) and fcc (111) surfaces of $4d$- and
$5d$-transition metals, which are also experimentally accessible,
e.g.~Fe/Ir(111)~\cite{AndrieuPRB1992,vBergmannPRL2006},
Fe/Ru(0001)~\cite{LiuPRB1990}, or
Fe/Pt(111)~\cite{RepettoPRB2006}.
We show that the nearest-neighbor exchange interaction, $J_1$, in the
Fe ML changes continuously
from antiferro- to ferromagnetic (FM) with filling of the
substrate $d$-band. Due to topological frustration on a triangular
lattice, AFM coupling for Fe on Re(0001) and Ru(0001) leads to a
N\'eel ground state with angles of 120$^\circ$ between adjacent
spins. This finding can explain unresolved experimental studies
reporting the absence of a ferromagnetic signal for
Fe/Ru(0001)~\cite{LiuPRB1990,BauerJPCM1999,AndrieuPRB1992,
TianSSC1991}.

For an Fe ML on substrates
such as Ru, Os, Rh, or Ir, $J_1$ is small and
interactions beyond nearest-neighbors or higher-order spin
interactions can be relevant. We exemplify this by studying
so-called multi-$\mathbf Q$ states, a superposition of symmetry
equivalent spin spirals,
which are degenerate in the
Heisenberg model but can gain energy, e.g., due to the presence of
biquadratic or four-spin interactions~\cite{KurzPRL2001}. In the
$4d$-TM substrates, significant magnetic moments are induced by
the Fe ML due to their high susceptibility.
For Fe/Rh(111), this is a crucial effect
which stabilizes an unexpected collinear ground-state of ferromagnetic
double-rows coupling antiferromagnetically along the $[11{\bar
2}]$-direction, a 2D analog of the antiferromagnetic
bilayer state in Fe films on Cu(001)~\cite{Asada1997}. For $5d$-TM
substrates such as Ir or Re, the large spin-orbit coupling may
cause a significant Dzyaloshinskii-Moriya interaction in the Fe ML
which can crucially affect the magnetic order~\cite{Bode2007,Ferriani2008},
making these systems a unique playground for complex magnetism.
%

We have determined the electronic and magnetic properties of 1 ML
Fe on the hexagonal hcp (0001) and fcc (111) surfaces of $4d$- and
isoelectronic $5d$-transition metals based on density-functional theory.
Calculations have been carried out in the generalized gradient
approximation (GGA) to the exchange-correlation
functional~\cite{ZhangPRL1998} using the full-potential linearized
augmented plane wave (FLAPW) method, as implemented in the {\sc
fleur} code~\cite{FLEUR,KurzPRB2004}.

The collinear magnetic states were investigated in systems modelled
by 6 or 7 layers of $4d$- or  $5d$-TM substrate with hcp or fcc
stacking covered by a pseudomorphic Fe monolayer on each side of the films.
We have used the experimental lattice constants which are very
close to the values obtained by GGA.
The structural relaxation of the Fe overlayer has been performed
for both fcc and hcp stacking. The non-collinear magnetic states
have been studied employing an asymmetric film consisting of four
substrate layers and an Fe monolayer on one side of the film at
the distance optimized for the collinear (FM or AFM) state of
lowest energy.
For Fe on Rh(111) we have used 6 substrate layers; we found that
adding two layers of substrate  did not influence the spin-spiral
dispersion by more than $6$~meV.
The spin spirals have been calculated
exploiting the generalized Bloch theorem~\cite{Herring66}. 
We have used about 100 basis functions
per atom for all calculations and at least 676 $\mathbf
k_\parallel$ points in the two-dimensional Brillouin zone (2D-BZ)
for the spin-spiral calculations, 48 $\mathbf k_\parallel$ points
in one quarter of the 2D-BZ for the $uudd$ configurations and 32
$\mathbf k_\parallel$ points in the 2D-BZ for the $3\mathbf
Q$-state requiring a surface unit-cell comprising 4 atoms.

\begin{figure}
\begin{center}
\centerline{\includegraphics[width=0.33\textwidth,angle=270]{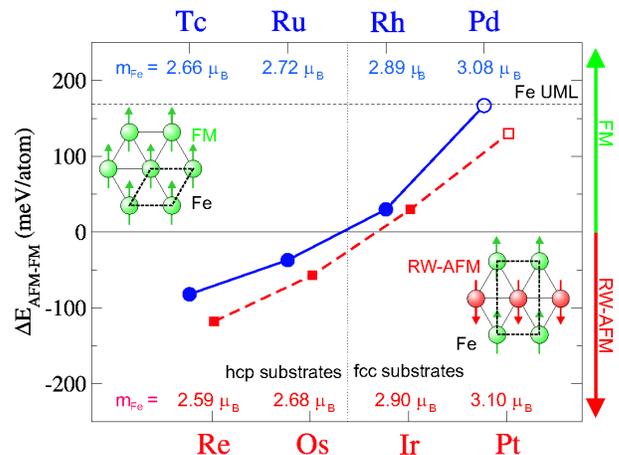}}
\caption{(color online)
Total-energy difference between the FM ($\Delta E>0$) and RW-AFM
($\Delta E<0$) configuration for Fe MLs on hcp (0001) and fcc
(111) surfaces of $4d$- and $5d$-TMs.
Closed and open  symbols
indicate a favorable hcp or fcc stacking of the Fe ML,
respectively. The magnetic moment of the Fe atoms, $m_{\rm Fe}$, is given.
$\Delta E_{\rm AFM-FM}$ is nearly constant for an unsupported
hexagonal Fe ML (UML) with the corresponding in-plane lattice
constants (dashed line).}
\label{fig:trend}
\end{center}
\vspace*{-1cm}
\end{figure}

In order to find the magnetic ground state,
we start by evaluating the total-energy difference between the FM
and the row-wise AFM (RW-AFM) configuration, Fig.~\ref{fig:trend},
considering hcp and fcc stackings of the monolayer.
Only for substrates at the end of the TM-series, Pd and Pt, the Fe
monolayer prefers fcc stacking. On all other substrates Fe prefers
an hcp stacking. Only on Rh and Ir the energy difference between
fcc and hcp stacking is sufficiently
small (9.0 and 7.6 meV/Fe-atom, respectively)
to suggest the experimental
observation of both
types~\cite{vBergmannPRL2006,vBergmannNJP2007}.

From Fig.~\ref{fig:trend} it can be concluded that
Fe on substrates from the center of the TM-series, Tc
and Re, exhibits a clear antiferromagnetic behavior and the RW-AFM
state has the lowest energy. Fe on substrates at the end of the
TM-series is ferromagnetic. In between we observe a gradual change
from a strongly AFM behavior to a strongly FM
one as function of the electron filling of the substrate.
It is argued that this change in the magnetic coupling results
from the $3d-4d$ and $3d-5d$ hybridization between the Fe ML and
the substrate, which is altered by the $d$-band filling. This
argument is supported by the fact (i) that the role of the
hybridization is also apparent from the monotonous variation of the
Fe magnetic moments as one moves through the TM-series and (ii)
the gradual change from FM to AFM coupling cannot be explained on
the basis of the changing in-plane lattice constant as the
comparison with unsupported MLs on the respective lattice
constants shows a rather stable ferromagnetic value of about
160~meV/Fe-atom, c.f.~Fig.~\ref{fig:trend}.

From the above results we can conclude that the antiferromagnetic
exchange interaction is strong and important for Fe on most of
these substrates except Pd and Pt on which Fe is clearly
ferromagnetic. The antiferromagnetic interaction on a triangular
lattice leads to the
frustration of magnetic interactions and is the origin of complex
magnetic states. Fe on Re or Tc exhibits strong antiferromagnetic
interactions as shown by the large energy gain when assuming a
RW-AFM state and the true ground state could be a
120$^\circ$-N\'eel state.
Due to the small energy difference between the FM and RW-AFM order
for Fe on Os, Ru, Rh and Ir many magnetic states have to be considered
as possible ground states. Therefore, we first focus on Fe MLs on Ru(0001) and on
Rh(111) as model systems of complex magnetism on a triangular
lattice~\cite{FeIr111}.

\begin{figure}
\begin{center}
\centerline{\includegraphics[width=0.45\textwidth,angle=0]{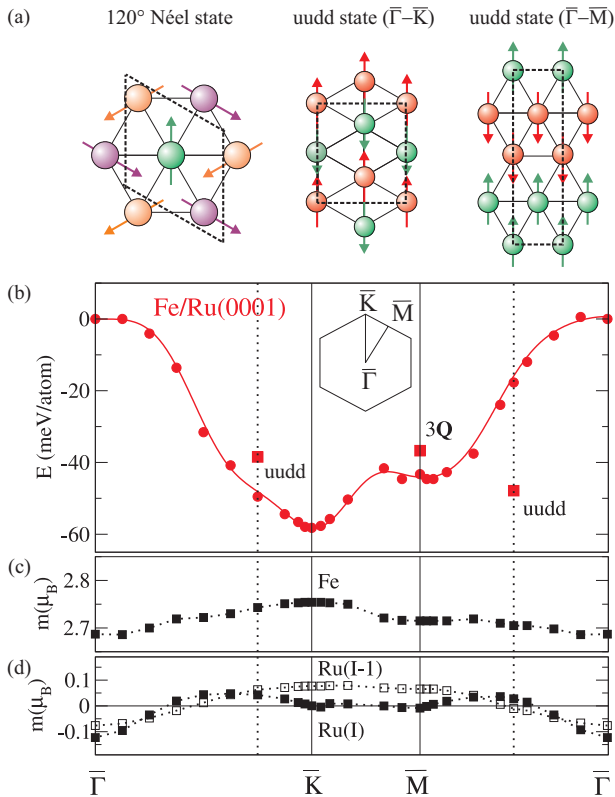}}
\caption{(color online) (a) N\'eel state and $uudd$-states
obtained from a superposition of two $90^\circ$~spin spirals
running either along $\overline{\Gamma {\rm K}}$ or
$\overline{\Gamma {\rm M}}$. (b) Total energy of spin spirals for
one hcp monolayer Fe on Ru(0001) (filled circles) along the high
symmetry directions of the 2D-BZ (see inset). Solid line denotes a
fit to the Heisenberg model up to $J_5$. The $3\mathbf Q$-
(downward triangle) and $uudd$-state (upward triangle) are
included for comparison. Magnetic moments of (c) Fe atoms and (d)
Ru interface atoms, Ru(I) (filled squares) and sub-interface
atoms, Ru(I-1) (open squares).} \label{FeRu}
\end{center}
\vspace{-1.cm}
\end{figure}

We study non-collinear magnetic structures by performing
spin-spiral calculations. Flat spin spirals are the general
solution of the classical Heisenberg model on a periodic lattice
$H=-\sum_{i<j}J_{ij}\:\hat{\mathbf{M}}_i\cdot\hat{\mathbf{M}}_j$,
where the exchange constants $J_{ij}$ determine the strength and the
type of coupling between local moments at sites $i$ and $j$
pointing along the unit vectors $\hat{\mathbf M}_i$ and
$\hat{\mathbf M}_j$, respectively. Spin spirals are characterized
by a wave vector $\mathbf q$ and the moment of an atom at site
$\mathbf R_i$ is given by $\mathbf M_i(\mathbf
R_i)=M\big(\cos(\mathbf q\cdot\mathbf R_i),\sin(\mathbf
q\cdot\mathbf R_i),0\big)$, where $M$ is the spin moment per atom.
By considering spin spirals along the high symmetry lines of the
2D-BZ we cover an important part of the magnetic phase space.
At high symmetry points, we find well-known magnetic states such as the FM
state at the $\overline{\Gamma}$-point, the RW-AFM state at the
$\overline{\mathrm{M}}$-point, and the $120^\circ$ N\'eel state
at the $\overline{\mathrm{K}}$~point, c.f.~Fig.~\ref{FeRu}(a).

The calculated total-energy dispersion, $E(\mathbf q)$, of spin
spirals for Fe/Ru(0001), Fig.~\ref{FeRu}(b), shows that the N\'eel
state, i.e.~for ${\mathbf q}=\overline{\rm{K}}$, is the
energetically most favorable magnetic configuration. The energy
gain is $58$~meV/Fe-atom with respect to the FM state and only
$13$~meV/Fe-atom with respect to the RW-AFM state. The magnetic
moment in the Fe ML, Fig.~\ref{FeRu}(c), depends only weakly on
$\mathbf q$.
Interestingly, the moments of the surface and subsurface Ru
layer are of similar magnitude and change their alignement
to the Fe moments, Fig.~\ref{FeRu}(c)-(d).

To interpret our calculations, we have mapped
the results onto a 2D Heisenberg model which allows us to
determine effective exchange constants $J_{ij}$ between moments of
the Fe atoms in the ML. Fig.~\ref{FeRu}(b) shows
that a good fit is obtained by including up to five nearest
neighbors. From the $J_{ij}$'s, given in
table~\ref{tab_Jij}, we conclude that the main effect of the
substrate is to change the nearest-neighbor exchange coupling,
$J_1$, from FM ($J_1>0$) for Rh and Ir to AFM ($J_1<0$) for Ru and Re.

For itinerant magnets such as iron, it is not {\it a priori} clear
that the Heisenberg model, which relies on localized magnetic
moments, can provide a good description.
The next higher-order terms beyond the Heisenberg model are the
biquadratic and four-spin interactions. These terms can lead to
energy contributions on the order of
$15$~meV/Fe-atom~\cite{KurzPRL2001} which is similar to the energy
difference between N\'eel and RW-AFM state of Fe/Ru(0001).

In order to find the magnitude of these interactions, we consider
linear combinations of spin spirals, the multi-$\mathbf Q$
states. The degeneracy of single- and multi-$\mathbf Q$ states
within the Heisenberg model is lifted
by higher-order
interactions. By calculating the energy differences between
suitable single-$\mathbf Q$
and multi-$\mathbf Q$
states, we can obtain values for the
nearest-neighbor biquadratic interaction, $B$, and four-spin
interaction, $K$.
One pair of such states are spin spirals at the $\overline{{\rm
M}}$-points of the 2D-BZ and the $3\mathbf Q$-state constructed
from the three independent $\overline{{\rm
M}}$-points~\cite{KurzPRL2001}. As a second pair we choose the
spin spiral at $\mathbf Q_{3\overline{\mathrm K}/4}$ and a
superposition of two such spirals with opposite rotation sense,
a so-called $uudd$-state (middle panel of Fig.~\ref{FeRu}(a)).
The total-energy differences are
\begin{eqnarray}
E_{3Q}-E_{\mathrm{RW-AFM}}&=&(16/3)\{2K+B\}\\
E_{uudd}-E_{3\overline{\mathrm K}/4}&=&4\{2K-B\}
\end{eqnarray}
marked by squares in Fig.~\ref{FeRu}(b) and
the obtained constants $B$ and $K$ are given in
table~\ref{tab_Jij}.

The large energy differences between the multi-$\mathbf Q$ and
single-$\mathbf Q$
states of 7 to 11~meV/Fe-atom
found for Fe/Ru(0001) clearly demonstrate its itinerant
character. The extracted values for the biquadratic and four-spin
interaction are of the order of the exchange constants beyond
nearest neighbors and cannot be neglected in finding the ground
state. Higher-order interactions play a similar
role for other substrates, as shown in table~\ref{tab_Jij}.

The Fe ML also induces considerable magnetic moments in the Ru
substrate due to its high susceptibility. The induced magnetic
moments of Ru depend on the magnetic structure in the Fe ML and
their size contributes to the total energy of the system. This
effect can be dramatic as seen
by comparing the $uudd$-state along
$\overline{\Gamma}-\overline{\rm M}$ and
$\overline{\Gamma}-\overline{\rm K}$,  c.f.~right panel of
Fig.~\ref{FeRu}(a), to  the corresponding
single-$\mathbf Q$
states. The energy difference due to biquadratic and four-spin
interactions should be the same in both cases, which is clearly
not the case. This finding can be traced back to the difference in
the induced magnetic moments in the substrate.

\begin{table}
\caption{\label{tab_Jij}\small{Heisenberg exchange constants for
the hcp Fe ML on different substrates obtained by fitting the
total-energy dispersion along $\overline{\Gamma}-\overline{\mathrm
K}-\overline{\mathrm M}$ and higher order Heisenberg terms.}}
\begin{ruledtabular}
\begin{tabular*}{\hsize}{l@{\extracolsep{0ptplus1fil}}cccccccc}
(meV)&$J_1$&$J_2$&$J_3$&$J_4$&$B$&$K$\\\hline
\colrule
Fe/Re(0001) & $-$14.5 & $-$0.5 & $-$5.4 & $-$0.5 &  3.1 & 1.8 \\
Fe/Ru(0001) &  $-$6.4 &  0.7 & $-$0.3 &  0.4 & $-$0.6 & 1.1 \\
Fe/Rh(111)  &   3.8 & $-$0.6 & $-$1.0 &  0.3 & $-$1.9 & 0.6 \\
Fe/Ir(111)  &   4.2 & $-$0.8 &  0.3 &  0.2 &  0.7 & 0.4
\end{tabular*}
\end{ruledtabular}
\end{table}

\begin{figure}
\begin{center}
\centerline{\includegraphics[width=0.45\textwidth,angle=0]{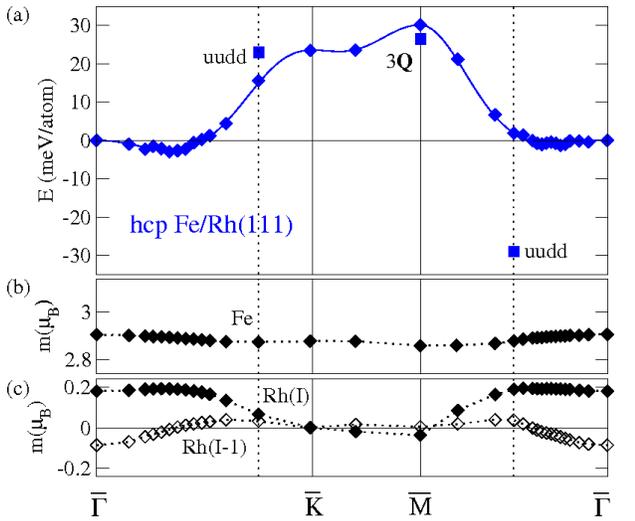}}
\caption{(color online) (a) Total energy of one hcp monolayer Fe
on Rh(111) for spin-spirals along high symmetry lines
(c.f.~Fig.~\ref{FeRu}) and multi-$\mathbf Q$
states. Solid line denotes a fit to the Heisenberg model up to
$J_5$. Magnetic moments of the Fe atoms and the Rh
interface atoms (Rh(I), open symbols) and sub-interface atoms
(Rh(I-1), filled symbols) are shown in panels (b) and (c), respectively.} \label{FeRh}
\end{center}
\vspace{-1.cm}
\end{figure}

We now turn to Fe on Rh(111) for which the FM state was slightly
favorable among the collinear states considered in
Fig.~\ref{fig:trend}. Our spin-spiral calculations, however,
indicate a non-collinear ground state with a spiral vector
$\mathbf q$ of about $0.225\times \frac{2\pi}{a}$ along
$\overline{\Gamma}-\overline{\rm K}$, about $2.5$ meV/Fe-atom below
the FM state (Fig.~\ref{FeRh}). From the fitting to the Heisenberg
model, c.f.~table~\ref{tab_Jij}, we indeed find a very small FM
nearest-neighbor exchange constant while second- and third-nearest
neighbors prefer an AFM alignment of a magnitude comparable to the
FM exchange.

Including the multi-$\mathbf Q$
states in our search for the ground-state results in the
$uudd$-solution ($\overline{\Gamma}-\overline{\rm M}$) as the most
favorable configuration with a large energy gain of 29.0~meV/Fe-atom
with
respect to the single-$\mathbf Q$ state.
However, the real magnetic ground-state might be even more complex
due to the competing interactions involved making Fe/Rh(111) a
truly intriguing system and a challenge for an experimental
investigation.

\begin{figure}
\begin{center}
\centerline{\includegraphics[width=0.4\textwidth,angle=0]{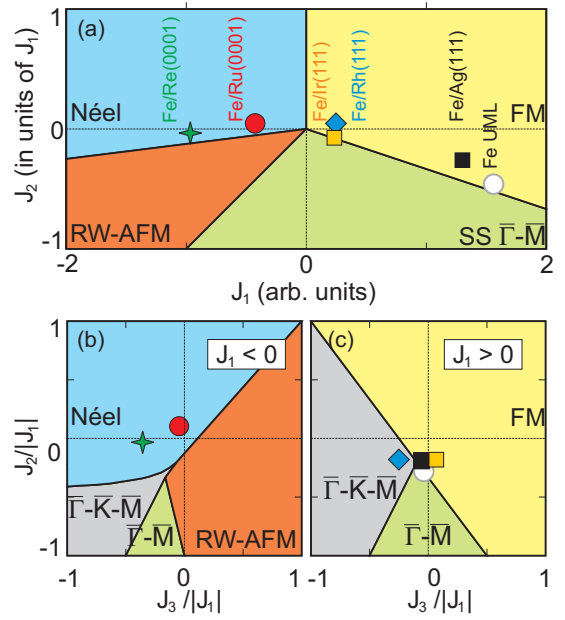}}
\caption{ (color online)
Phase diagrams of the classical Heisenberg model for a 2D
hexagonal lattice for states on the high symmetry lines. (a) $J_1$-$J_2$
plane for $J_3=0$. $J_2$-$J_3$ plane for (b) $J_1>0$ and (c) $J_1<0$.
(Filled symbols denote values obtained from
fits for the Fe ML on $4d$- and $5d$-TM substrates.)
} \label{phase}
\end{center}
\vspace*{-1cm}
\end{figure}

In the phase diagram of the 2D Heisenberg model, shown in
Fig.~\ref{phase}, we can provide a complete picture of the
substrates' impact on the Fe exchange coupling by including spin
spiral calculations for an Fe ML on Re(0001), Ir(111), and
Ag(111). In the $J_1$-$J_2$ plane of the diagram,
Fig.~\ref{phase}(a), we see that the $d$-band filling of the
substrate drives the system along the line $J_2 \approx 0$ from a
N\'eel configuration on Re and Ru to the FM solution on Ir and Rh.
For small $J_1$, we need to consider also the phase diagrams in the
$J_2$-$J_3$ plane showing the spin spiral minimum of Fe on Rh(111)
in the $\overline{\Gamma {\rm KM}}$-direction,
Fig.~\ref{phase}(c), and the $120^{\circ}$ N\'eel state of Re(0001),
Fig.~\ref{phase}(b). In contrast, the FM state is the most
favorable single-$\mathbf Q$
state for the hcp Fe ML on Ir(111)~\cite{FeIr111}.

In conclusion, we have proposed Fe MLs on hexagonal surfaces of
late $4d$- and $5d$-TMs as promising systems to study
experimentally the 
magnetic interactions in TMs, e.g.~by proving the influence of
spin interactions beyond the Heisenberg model. Alloying of the
substrate, e.g.~PtRu(0001)~\cite{Preprint_Behm2008},  may allow an
additional fine tuning of the degree of disorder in 2D systems
as proposed in Ref.~\cite{FerrianiPRL2007}.


Financial support of the Stifterverband f\"ur die Deutsche
Wissenschaft and the Interdisciplinary Nanoscience Center Hamburg
and the ESF EUROCORES Programme  SONS under Contract No. ERAS-CT-2003-980409
is gratefully acknowledged.


\end{document}